\documentclass[11pt,a4paper]{article}

\usepackage{fullpage}
\usepackage{amsmath}
\usepackage{hyperref}
\usepackage{graphicx}

\begin{document}

\renewcommand{\topfraction}{1} 
\renewcommand{\bottomfraction}{1}
\renewcommand{\floatpagefraction}{1}
\renewcommand{\textfraction}{0}


\newcommand{\ie}{i.e.,\ }
\newcommand{\eg}{e.g.,\ }

\newcommand{\const}{\operatorname{const.}} 


\newcommand{\rmd}{\,\mathrm{d}}

\newcommand{\Tr}{\operatorname{tr}}

\newcommand{\e}[1]{\operatorname{e}^{#1}}

\newcommand{\op}{\mathcal{O}}

\newcommand{\vev}[1]{\left\langle #1 \right\rangle}
\newcommand{\fvev}[1]{\langle #1 \rangle}
\newcommand{\jvev}[1]{\left\langle j\left| #1 \right| j \right\rangle}

\newcommand{\comb}[2]{\begin{pmatrix} #1\\#2\end{pmatrix}}

\newcommand{\Lag}{\mathcal{L}}
\newcommand{\Ham}{\mathcal{H}}

\newcommand{\Order}{\mathcal{O}}

\newcommand{\Mpl}{M_{\text{P}}}
\newcommand{\Lpl}{L_{\text{P}}}

\newcommand{\pois}[2]{\left\{#1,#2\right\}}
\newcommand{\dirac}[2]{\pois{#1}{#2}_D}
\newcommand{\commut}[2]{\left[#1,#2\right]}

\newcommand{\bx}{\mathbf{x}}
\newcommand{\by}{\mathbf{y}}
\newcommand{\bk}{\mathbf{k}}
\newcommand{\bv}{\mathbf{v}}

\begin{center}

{\Large \textbf{On the number of soft quanta in classical field configurations}}\\[2em]

\renewcommand{\thefootnote}{\fnsymbol{footnote}}
Wolfgang M{\"u}ck${}^{a,b}$\footnote[1]{wolfgang.mueck@na.infn.it}\\[2em]
\renewcommand{\thefootnote}{\arabic{footnote}}
${}^a$\emph{Dipartimento di Fisica, Universit\`a degli Studi di Napoli "Federico II"\\ Via Cintia, 80126 Napoli, Italy}\\[1em] 
${}^b$\emph{Istituto Nazionale di Fisica Nucleare, Sezione di Napoli\\ Via Cintia, 80126 Napoli, Italy}\\[1em]

\abstract{A crucial ingredient in the large-$N$ quantum portrait of black holes proposed by Dvali and Gomez is the estimate of the number of soft quanta that make up the classical gravitational field. It is argued here that the coherent state formalism provides a way to calculate that number directly. As a consequence, the average energy of the soft quanta is such that the typical geometric size of the field source can be roughly interpreted as their de~Broglie wavelength. The calculation is done for the electromagnetic field and for Newtonian gravity, and it is argued that the number of soft quanta should be unchanged in General Relativity due to the long range nature of gravity.}

\end{center}

\paragraph{Introduction}
Recently, Dvali and Gomez proposed the phenomenologically fascinating picture of a black hole as a system of a very large number, $N$, of gravitons at the verge of a quantum phase transition to a Bose Einstein condensate (BEC) \cite{Dvali:2011aa}. The number $N$ of soft quanta was introduced as a measure of classicality in the more general context of classicalizing theories, of which gravity is considered the most striking and efficient example \cite{Dvali:2011th}. For a body of mass $M$, $N$ is estimated as 
\begin{equation}
\label{N.estimate}
	 N = \frac{M^2}{\Mpl^2}~,
\end{equation}
possibly up to a numerical factor, where $\Mpl$ is the Planck mass. Although $N$ is huge for macroscopic bodies, semi-classical physics may break down when collective quantum effects become important, \ie when the condensation of the soft quanta sets in, which is allegedly the case for black holes. 
In the absence of a theory that would describe the condensation of gravitons, toy models are useful to test the main features of this picture. To name a few, the importance of long-range quantum effects in interacting boson systems undergoing a quantum phase transition has been studied by Flassig, Pritzel and Wintergerst \cite{Flassig:2012re}. The results of Berkhahn, M{\"u}ller, Niedermann and Schneider \cite{Berkhahn:2013woa} can been interpreted as further evidence in favour of a quantum phase transition in black holes. Casadio and Orlandi \cite{Casadio:2013hja} have presented a simple quantum-mechanical toy model for the wave function of the condensate, which incorporates the main features of the BEC picture, such as marginal binding of the gravitons. Finally, information retrieval from black holes in the context of String Theory, which should be sensitive to the $1/N$ quantum hair present in the BEC picture, has been studied by Veneziano \cite{Veneziano:2012yj}. For more recent contributions on the subject, see \cite{Dvali:2013vxa, Dvali:2013lva}.

If one accepts \eqref{N.estimate}, a very simple argument indicating that black hole formation corresponds to some sort of phase transition may be formulated as follows. Consider an object of mass $M$ and characteristic size $R$. The average energy of the $N$ gravitons can be estimated by identifying $R$ with their de~Broglie wavelength. Hence, their total energy will be $N\hbar/R=M^2G/R$.\footnote{Throughout this letter, Planck's constant and the gravitational constant are expressed in terms of the Planck length and mass, $\hbar = \Mpl\Lpl$, $G=\Lpl/\Mpl$.} For $R$ of about the size of the gravitational radius, $r_g=2GM$, the soft gravitons will carry about all of the system's energy $M$, which implies that $R$ cannot be lowered further without emitting some energy. Such an emission does occur and is nicely described in the BEC picture as the depletion of the condensate due to quantum fluctuations \cite{Dvali:2011aa}. 

It is evident that \eqref{N.estimate} plays the key role in this argument. Similar estimates are needed for arguments in favour of the self-completeness of classicalizing theories, where field configurations made up of a large number of soft quanta, so-called \emph{classicalons}, protect the theory from ultraviolet divergences. To the best of my knowledge, a calculation of \eqref{N.estimate} going beyond the estimate given in \cite{Dvali:2011th}, has not appeared in the literature. In what follows, I propose that, interpreting a classical field configuration in terms of coherent states in a quantum theory, the number of soft quanta can be directly calculated. Moreover, it is shown that the average energy of the quanta supports the idea of identifying the typical size of the object as a de~Broglie wavelength of soft quanta. For simplicity, the case of a static electric field around a charged ball is considered. The equivalence between the laws of electrostatics and Newtonian gravity then leads to \eqref{N.estimate}. But let us start with a short review of coherent states.

\paragraph{Coherent States}
It is well known that coherent states, which are eigenstates of field annihilation operators, are those states that most closely correspond to classical field configurations. Conversely, let $\hat{\phi}(x)$ and $\hat{\phi}^\dagger(x)$ be some field annihilation and creation operators, respectively.\footnote{They satisfy $\left[\hat{\phi}(x),\hat{\phi}^\dagger(y)\right] =\delta^3(x-y)$.} One may define the coherent state $|\phi\rangle$ corresponding to the classical field configuration $\phi(x)$ by 
\begin{equation}
\label{coh.state}
	\hat{\phi}(x) |\phi\rangle = \phi(x)|\phi\rangle~.
\end{equation}
Explicitly, $|\phi\rangle$ is given by
\begin{equation}
\label{coh.state.expl}
	|\phi\rangle = \e{-\frac12 N} \e{\int\rmd^3x\, \phi(x)\hat{\phi}^\dagger(x)} |0\rangle~,
\end{equation}
where $|0\rangle$ is the vacuum and $N$ coincides with the expectation value of the number operator $\hat{N}=\int \rmd^3x\,\hat{\phi}^\dagger(x)\hat{\phi}(x)$ in the coherent state $|\phi\rangle$,
\begin{equation}
\label{N}
	N = \langle \phi |\hat{N}|\phi\rangle = \int \rmd^3x\,|\phi(x)|^2~.
\end{equation}
Notice that, in order for the units to match, $\phi$ must have units $L^{-3/2}$. In order to make quantum properties evident, I shall distinguish between mass and inverse length units by keeping $\hbar$ explicit.
An interesting quantity is the relative particle number uncertainty, $(\Delta N)/N$, where $(\Delta N)^2 = \langle \phi | \hat{N}^2 -N^2 | \phi\rangle$, which is found to be  $(\Delta N)/N= 1/\sqrt{N}$. Therefore, the particle number is a good observable in a coherent state only if it is large.

\paragraph{Static Electric Field}
Consider the static electric field around a spherical shell of radius $R$ carrying a charge $q$. In spherical coordinates, the electrostatic potential\footnote{In order to avoid confusion with the classical field $\phi$ introduced earlier, $A_0$, the time-component of the 4-vector $A_\mu$, is used to denote the potential.} is given by 
\begin{equation}
\label{A}
	A_0 = \begin{cases} \frac{q}{4\pi r} \qquad & \text{for $r>R$,}\\
				 \frac{q}{4\pi R} \qquad & \text{for $r\leq R$.}
		  \end{cases} 
\end{equation}
The field carries canonical units such that $[A_0]= (M/L)^{1/2}$ and $[q]=(ML)^{1/2}$. In order to interpret the classical field \eqref{A} as a coherent state of (time-like) photons, we need to multiply it by a constant of dimension $M^{-1/2}L^{-1}$. However, in contrast to interactions that allow for classicalization \cite{Dvali:2010jz}, electrodynamics does not have a coupling that would allow the construction of such a constant in combination with $\hbar$, because $[q^2]=[\hbar]$. Hence, an auxiliary scale must be introduced, which we can identify as a tiny photon mass $\mu$, the same that is introduced in Quantum Electrodynamics to pragmatically deal with the infrared problem. Consequently, the long-distance fall-off in \eqref{A} should be corrected by a Yukawa potential, but for simplicity we shall keep the $1/r$ behaviour and simply use a poor-man's regularization cutting off divergent integrals at $r=\hbar/\mu$. Following this simpler procedure will yield the correct result up to a possible numerical factor of order unity, which is sufficient for our purposes.\footnote{An exact treatment of the quantized electromagnetic field with non-zero mass $\mu$ supports this procedure \cite{Mueck2013:xxx}.}
Then, up to a numerical constant, the field $\phi$ with number density units is identified as
\begin{equation}
\label{phi.A.def}
		\phi = \frac{\sqrt{\mu}}{\hbar} A_0~.
\end{equation}
Using \eqref{A} and \eqref{phi.A.def}, the number of soft field quanta \eqref{N} is immediately found as
\begin{equation}
\label{N.A}
	N = \frac{q^2}{4\pi\hbar} \left( 1-\frac{2\mu R}{3\hbar} \right) \approx 	\frac{q^2}{4\pi\hbar}~.
\end{equation} 
Expressing the charge $q$ in terms of the elementary charge $e$ by $q=ne$, \eqref{N.A} takes the suggestive form $N=\alpha n^2$, where $\alpha$ is the fine structure constant. 

A number of comments are in order. First, it is evident that the finite result \eqref{N.A} stems from the term, in which the fictitious photon mass has cancelled against the large distance cut-off. Second, if the total charge is non-zero, the geometric distribution of the charge does not influence the result, as any term involving geometric distances is necessarily multiplied by $\mu$ and can be neglected. However, a field configuration corresponding to zero total charge would give rise to a tiny (proporional to $\mu$) photon number. As observed above, this would imply a huge (proportional to $\mu^{-1/2}$) photon number uncertainty, so that the photon number is not a good observable in such a classical state. Third, one may wonder how the definition \eqref{phi.A.def}, which relates the electric potential to a number density, compares to the treatment of physical photons in laboratory experiments. At first sight, the number of photons in any finite volume would appear to be vanishingly small, but this is incorrect, because real photons are relativistic particles. Again, consider a photon as a particle with a tiny rest mass $\mu$. If its energy in the laboratory frame is $\hbar \omega$, its boost factor compared to the rest frame is $\gamma=\hbar\omega/\mu$. Hence, a rest frame number density, as suggested by \eqref{phi.A.def}, $\rho_0\sim \mu A^2/\hbar^2$ would lead to a measurable density in the laboratory frame $\rho=\gamma\rho_0\sim \omega A^2/\hbar$. Here, $A^2$ stands for a relativistic invariant of the field. The photon mass has again cancelled. 

The energy of the electric field \eqref{A} is $E=q^2/(8\pi R)$, where small, $\mu$-dependent terms have been neglected. Hence, we conclude that the average energy of the soft photons is 
\begin{equation}
\label{E.average}
	m = \frac{E}{N} = \frac{\hbar}{2R}~,
\end{equation} 
which supports the interpretation of the size $R$ as a typical de~Broglie wavelength. Note that the precise numerical factor depends on the geometrical distribution of the charge, because $E$ depends on it. For example, if we consider a charged ball with homogeneous charge density instead of a charged  spherical shell, \eqref{E.average} would be multiplied by a factor of $6/5$.

\paragraph{Newtonian Gravity}
The results obtained for electrostatic fields can be translated directly to Newtonian gravity. The simplest way to do this is to introduce a ``gravitational charge'', $q_M=\sqrt{4\pi G}M$, such that Newton's law of gravitation and Coulomb's law become identical. Thus, the number of soft gravitons making up the gravitational field around a mass $M$ follows from \eqref{N.A},
\begin{equation}
\label{N.grav}
	N =\frac{4\pi GM^2}{4\pi \hbar} = \frac{M^2}{\Mpl^2}~,
\end{equation}
which is precisely \eqref{N.estimate}. As before, a tiny graviton mass must be introduced to cure the long distance singularity and to convert the field to number density units. This is related to the fact that gravity is a long range force, which renders $N$ independent of the geometrical distribution of the mass. 
The average energy of the gravitons is obtained by dividing the gravitational energy by $N$. This yields again \eqref{E.average} times a numerical factor that depends on the distribution of the mass. For example, for a ball of constant mass density (such as a marble), one gets $m=\frac{3\hbar}{5R}$. The conclusion is again that the characteristic size of the mass distribution is a good estimate for the de~Broglie wavelength of the soft gravitons.

One may argue that, in contrast to electromagnetism, gravity does not necessitate the introduction of a new mass scale $\mu$ to get the units right, because  $G$ and $\hbar$ provide the Planck mass, $\Mpl=\sqrt{\hbar/G}$. Of course, $\Mpl$ is too large to be used as a graviton mass, so that it would not be possible to cure the infrared divergence. One possible point of view, which we pragmatically adopt, is that $G$ is needed to convert the mass and the gravitational potential to canonical charge and field units, after which it disappears from the field equation of Newtonian gravity. Hence, we are led to use the same procedure as for the electric field. 

\paragraph{Conclusions}
It has been shown that, for the cases of a static electric field and for Newtonian gravity, the coherent state formalism can be used to calculate the number of soft quanta inherent in the classical field configurations. As a consequence, the average energy of the quanta corresponds to a typical de~Broglie wavelength of about the same size as the geometric extension of the fields' sources. This should be contrasted with the argument presented in \cite{Dvali:2011th}, where the latter statement was assumed in order to derive an estimate for $N$ for a \emph{classicalon} configuration. Electrodynamics is not a classicalizing theory, so there are no classicalon configurations. This simply implies that, no matter the geometric distribution of the charges, the $N$ soft photons cannot undergo a phase transition into a BEC. The same is true for the soft gravitons of Newtonian gravity, but not of General Relativity. It would be interesting to extend the above argumentation to General Relativity or other theories of gravity. In order to do so, one must find a suitable definition of the gravitational field that can be interpreted as the configuration defining a coherent state, and one must be able to identify the gravitational energy (not the Komar mass, which measures the total energy). Based on our results, in particular the fact that the number of soft quanta is due to the long-range nature of the forces, one can reasonably expect that \eqref{N.grav} remains unchanged. This is because, far enough from the matter sources, say for $r>R$, gravity is weak and the Newtonian approximation may be used. As argued above, this is sufficient to derive \eqref{N.grav}, up to corrections of order $\mu R/\hbar$, which can be neglected. What will be different, of course, is the expression for the gravitational energy, which depends on the mass distribution and must be such that it cannot exceed the Komar mass. If the BEC picture of black holes is correct, the coherent state should become the ground state. These interesting issues are left for the future.

\bibliographystyle{JHEP}
\bibliography{classicalization}
\end{document}